\documentstyle{article}
\newcommand{\beq}{\begin{equation}}
\newcommand{\eeq}{\end{equation}}
\newcommand{\beqa}{\begin{eqnarray}}
\newcommand{\eeqa}{\end{eqnarray}}

\def\bra#1{\langle #1 |} 
\def\ket#1{| #1 \rangle}

\def\half{\frac{1}{2}}
\def\F{{\cal F}}

\def\up{\uparrow}
\def\down{\downarrow}
\def\lefta{\leftarrow}
\def\righta{\rightarrow}

\def\vbasis{$\bf \updownarrow$ }
\def\hbasis{$\bf \leftrightarrow$ }

\title{Quantum Cloning, Eavesdropping and Bell's inequality}
\author{
N. Gisin and B. Huttner \\
Group of Applied Physics\\University of Geneva\\
20 Rue de l'Ecole de Medecine\\
CH 1211 Geneva 4\\ Switzerland}

\date{20/11/1996}

\begin{document}
\input{epsf}

\maketitle

\begin{abstract}
We analyze various eavesdropping strategies on a quantum cryptographic
channel. We present the optimal strategy for an eavesdropper restricted to
a two-dimensional probe, interacting on-line with each transmitted signal.
The link between safety of the transmission and the violation of  Bell's
inequality is discussed. We also use a quantum copying machine for
eavesdropping and for broadcasting quantum information. 
\end{abstract}

\section{Introduction}\label{int}
How good can a quantum ``photocopy machine'' be?  How many pairs of people
can use such a machine on a given 2-particle source and still 
violate Bell's inequality? How much information can an
eavedroper read out of a quantum cryptography channel for a given BER
(bit error rate)? Is quantum privacy amplification
intrinsically more powerful than its classical analog, or is it only
provably secure? How much are the above questions related to each other?

This letter plays with the above questions and provides several answers
that in turn raise more questions. The general motivation stems from
quantum cryptography, but the
problematic is more general: what can one do with quantum information that
can not be done with classical information? Indeed, quantum information
processing calls for original applications, not just mimicking classical
applications in a more efficient way. Shor's factorization
algorithm~\cite{shor}, for
example, is certainly not the end of the adventure, but a brilliant step
that
calls for more (realistic) imagination.

The paradigm of this letter is a source of entangled EPR particles,
assumed to be spin half particles, now known as qubits, which are shared
between two users, known as Alice and Bob. One particle goes directly to
Alice, while the other one may be modified on its way to Bob by a
malevolent eavesdropper, known as Eve. Alice and Bob may now use various
setups, either to distribute a secret key, or to test ``the
inequality''~\footnote{When John Bell
was talking of his inequality, he would just say ``the inequality''. We
shall adopt the same prescription in this work}~\cite{bell}. The choice of
setups on
both sides is summarized in Fig.~\ref{setup}. In the so-called BB84 or
4-states protocol for quantum cryptography~\cite{BB84}, Alice measures her
particles,
either in the
vertical basis: \vbasis,  or in the horizontal one: \hbasis. 
She chooses her basis at her
free will (or using a ``true'' random generator), independently of all
other
players in the protocol. Equivalently, we could assume that Alice prepares
particles in either basis and send them to Bob. This would be entirely
equivalent for our purposes.  
Bob does precisely the same as Alice (but making independent choices),
thus measuring the particles in either \vbasis or \hbasis bases.
Alternatively, Alice and Bob may choose to use their particles to
test the inequality. In this case, Bob simply rotates his reference frame
by 45 degrees. Finally, they may use the Ekert protocol for quantum
cryptography~\cite{ekert}, and choose  a combination of both setups, each
of them now
choosing between three possible bases. In this case, when they both use
the same basis, as in the BB84 protocol, they shall use the data to obtain
a key, while when they use different bases, they use the data to test the
inequality. 

To these ideal protocols, we shall add Eve, the malevolent
eavesdropper who would like to hear what Alice has to say to Bob. Her aim
is to obtain as much information as possible on the key exchanged between
Alice and Bob, while creating as little disturbance as possible. 
Actually Eve has the most interesting position in this game, as, in
principle, she is allowed to
do everything, being only limited by the laws of physics (and the size of
the Universe~\cite{IanCopenhagen}).
However, in order to keep the problem manageable, we shall impose two
conditions:
\begin{enumerate}
\item  She interacts with only one qubit at a time. 
\item  Her probe can be described by a two-dimensional Hilbert space (i.e.
it is also a qubit).
\end{enumerate}
These two restrictions do limit the validity of our approach, but they are
also the only realistic ones experimentally. Indeed, it is becoming
possible to make two qubits interact with one another~\cite{experiments},
but more
complicated systems are still quite a way off. A more general case, with
an auxiliary system  described by a four-dimensional
Hilbert space, was recently described by Bu\v{z}ek and
Hillery~\cite{buzek},
who devised a
universal quantum-copying machine (UQCM). We shall compare this machine
with a simpler two-dimensional one, and show that the UQCM is only
marginally better. Moreover, Fuchs and Peres~\cite{fuchs} have recently
shown that, as
long as the initial state of Eve's probe is a pure state, there is no need
to go beyond a four-dimensional space. They have also shown, although only
numerically, that the optimal detection method for a two-states system is
obtained with a two-dimensional probe only. 

\begin{figure}[ht]
\hspace*{1cm}\epsfbox{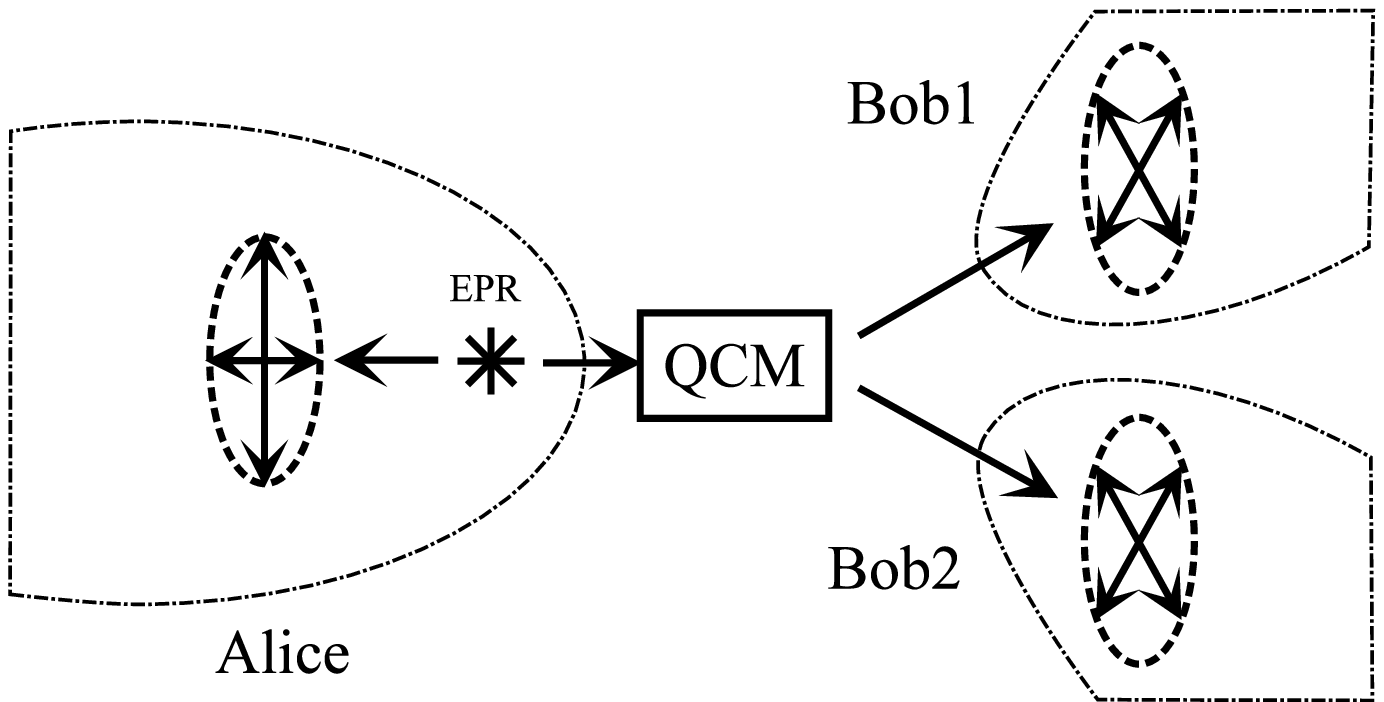}
\caption{Broadcasting quantum information}
Quantum cloning machines (QCM) allow to broadcast quantum
information, i.e. share it between various users,  at the cost of reducing
the fidelity of the channels. 
\label{QCM}
\end{figure}

In Section~\ref{Mgamma} we find the optimal strategy for Eve using a
two-dimensional probe.  This strategy is better than the
standard   ``intercept-resend'' strategy. It is also better than a
strategy that has
recently been published~\cite{lutkenhaus}, and was refered to as
``optimal''. The reason is that~\cite{lutkenhaus} is restricted to on-line
measurements: Eve has to perform her measurement immediately. Here, as we
introduce an explicit probe, we also allow Eve to delay her measurement on
her probe till Alice and Bob announce publicly the bases they used. We
find also that,
using this strategy, Eve can get reliable enough data to violate the
inequality
while Bob's data are still good enough to also violate the inequality!
Finally, the result of this Section prove that quantum privacy
amplification~\cite{qpa} is fundamentally
more efficient than classical error correction and privacy amplification
in the sense that for high enough BER, the latter is no longer possible,
whereas the former is still efficient.
In Section~\ref{Qclone} a pretty good quantum-copying machine (PGQCM)
machine is presented. The machine itself is  classical: only the original
qubit and the clone
are quantum, each represented by a 2-D Hilbert space. This pretty good
machine is then compared to Bu\v{z}ek and Hillery~\cite{buzek} UQCM. The
latter requires a
quantum machine, albeit the machine can also
be described by a 2-D Hilbert space. We lend these two machines to Eve 
and see that using either one provides her with more information than the
standard intercept-resend strategy, but less than the strategy studied in
Section~\ref{Mgamma}. Amusingly, we note that using such copy
machine Eve can send the (perturbed) original qubit to $\rm Bob_1$ and the
(poor)
copy to some $\rm Bob_2$, both Bobs violating Bell's inequalities (with
respect to Alice data)
or both Bobs establishing secret crypto keys with Alice.
This provides a new way to broadcast quantum information, as depicted in
Fig.~\ref{QCM}.

\section{Eavedropping with a 2-D probe}\label{Mgamma}
\begin{figure}[ht]
\hspace*{1cm}\epsfbox{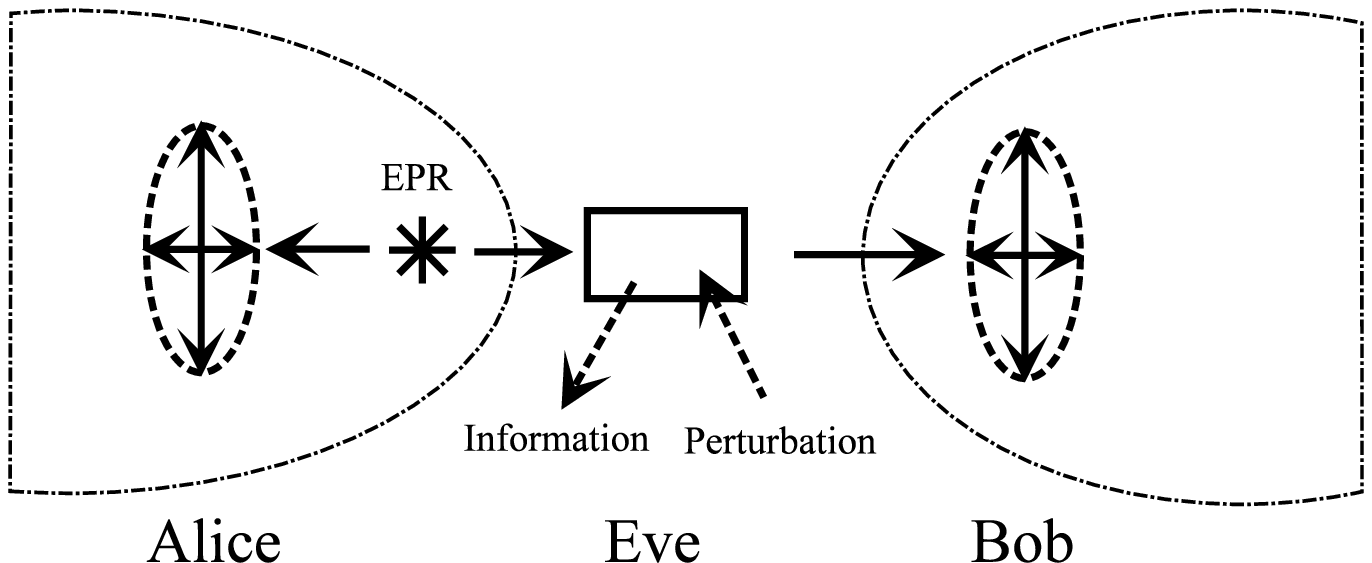}
\caption{Eavesdropping on a quantum channel}
Eve extracts information out of the quantum channel between
Alice and Bob at the cost of introducing noise into that channel.
\label{setting}
\end{figure}

\begin{figure}[ht]
\hspace*{5cm}\epsfbox{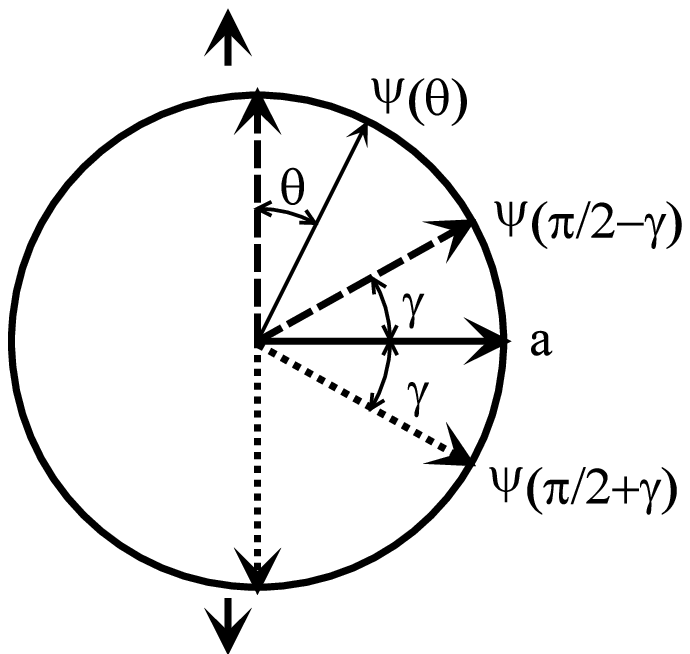}
\caption{}
State of the ancilla after the interaction corresponding to
measurements of intensity
$\gamma$ of an $\up$ state (dashed line,$\psi(\pi/2-\gamma)$) and of a
$\down$ state (dotted line,$\psi(\pi/2+\gamma)$). {\bf a} corresponds to
the symmetry breaking direction of the measurement.
\label{ancilla}
\end{figure}

In this Section we analyze and optimize the following eavedroping strategy
for Eve. The general setting is depicted in Fig.~\ref{setting}. When Alice
sends a qubit $\ket{\psi}$ to Bob, Eve lets a second qubit (often called
the
probe or the
ancilla)
interact with $\ket{\psi}$. Initially Eve's probe is in a know state 
$\ket 0$ and
the joint state of the unknown qubit and the probe is a product state
$\ket\psi\otimes\ket 0$. This product state undergoes some unitary
evolution after which the unknown qubit is forwarded to Bob who does his
standard measurement, irrespective of Eve's strategy. Eve may either
measure her probe immediately, as in the following measurement of
intensity $\gamma$, or keep her probe
until Alice and Bob reveal the bases used to encode this bit, and then
measure the probe and gains information about the corresponding  bit of
the cryptographic key. Of course Eve does not need to use always the same
unitary evolution. In this
way she can for instance restore the symmetry that some evolution may
break. She can also decide to reduce the perturbation, at the cost of
reducing
in the same ratio her information gain, by probing only a fraction of the
unknown qubits (this amounts to choosing the identity as unitary
evolution).
We shall only consider input states of the form
$\ket{\psi(\theta)}=\cos(\theta/2)\ket\up
+sin(\theta/2)\ket\down$. On the Poincar\'e (or Bloch) sphere these states
are represented by the points of a large circle that passes through the
antipodic points representing $\ket \up$ and $\ket \down$, and $\theta$ is
the angle
between the directions defined by the points representing $\ket \up$ and
$\ket \psi$ (Fig.~\ref{ancilla}).
In this way only real numbers need to be used. The unitary evolution is
thus defined by the following
real parameters $a_j$ and $b_j$:
\beqa
&\ket\up\otimes\ket 0& \rightarrow a_1\ket{\up\up}+a_2\ket{\up\down}+
        a_3\ket{\down\up}+a_4\ket{\down\down} \\ \nonumber
&\ket\down\otimes\ket 0& \rightarrow b_1\ket{\up\up}+b_2\ket{\up\down}+
        b_3\ket{\down\up}+b_4\ket{\down\down}
\label{Uab}
\eeqa
where unitarity implies $\sum_j |a_j|^2=\sum_j |b_j|^2=1$ and $\sum_j
a_j^* b_j=0$.

\underline{Assuming an $\up \down$ symmetry}, one has
$b_j=a_{5-j}$. Accordingly the general unitary transformation is
determined
by only 2 parameters that can be chosen as follows:
\beqa
&a_1&=\cos\alpha\cos\beta \hspace{1cm}
a_2=\cos\alpha\sin\beta \\ \nonumber
&a_3&=\sin\alpha\cos\beta \hspace{1cm}
a_4=-\sin\alpha\sin\beta 
\label{alphabeta}
\eeqa
For example the standard Von Neumann measurements correspond to $\alpha=
\beta=0$. This example can be generalized to another interesting case:
$\alpha=0$ and $\cos(\beta)^2=\frac{1+\sin\gamma}{2}$; $\gamma$ ranges
between 0 and
$\pi/2$. We call this {\it measurements of intensity} $\gamma$, as
$\gamma$ parametrizes the amount of information that Eve may obtain from
her probe. We shall
see below that, if we restrict Eve to a 2-D probe, this kind of
measurement optimizes Eve's information gain 
at low BERs, and is practically indistinguishable
from the optimum up to a BER of about 15\%. 
First, let us rewrite the measurement of
intensity $\gamma$ as follows, see Fig.~\ref{ancilla}:
\beqa
&\ket\up\otimes\ket 0& \rightarrow
\ket\up\otimes\ket{\psi(\frac{\pi}{2}-\gamma)} \\ \nonumber
&\ket\down\otimes\ket 0& \rightarrow
\ket\down\otimes\ket{\psi(\frac{\pi}{2}+\gamma)}
\label{mesIgamma}
\eeqa
When Alice sends state $\ket{\psi(\theta)}$, which corresponds to the
following density matrix:
\beq
\label{rhoalice}
\rho_{\rm Alice}=\half\pmatrix{1+\cos\theta & \sin\theta \cr
\sin\theta & 1-\cos\theta} \; ,
\eeq
the states at the disposal of
Bob and Eve, obtained by tracing out the other's qubit, read:
\beq
\label{rhobob}
\rho_{\rm Bob}=\half\pmatrix{1+\cos\theta & \cos\gamma\sin\theta \cr
\cos\gamma\sin\theta & 1-\cos\theta} 
\eeq
\vspace{3mm}
\beq
\label{rhoeve}
\rho_{\rm Eve}=\half\pmatrix{1+\sin\gamma\cos\theta & \cos\gamma \cr
\cos\gamma & 1-\sin\gamma\cos\theta} 
\eeq
For $\gamma=0$ Eve does not introduce any errors in the transmission (the
density matrix of Bob is unchanged), but  does not gain any information
either (her density matrix becomes independent of the initial state). For
positive
$\gamma$ some information about $\theta$ can be gained by measuring Eve's
probe and for $\gamma$ approaching $\pi/2$ this information is the optimum
Von Neumann scheme (from the information point of view). 
Simultaneously, Eve's interception introduces a 
perturbation on the initial state $\psi(\theta)$, as shown by
Eqs.~(\ref{rhoalice}) and~(\ref{rhobob}). We see that when Alice and Bob
use the \vbasis basis ($\theta = 0$ or $\theta=\pi$), Eve introduces no
perturbation: the state received by Bob is unchanged. However, she gains
information. For example, the
a posteriori probability of having input state $\ket \up$,  once she has
found her probe in the $\ket \up$
state is:
\beq P(\psi=\up |{\rm probe}= \up) = \frac{1+\sin(\gamma)}{2}\; . 
\eeq
On the other
hand, when Alice
and Bob choose the \hbasis basis ($\theta = \pm \pi/2$ ), Eve introduces
errors, but gains
no information at all (same $\rho_{\rm Eve}$ for both inputs). Hence Eve
gains nothing by
waiting
until she knows the bases: she can measure her probe immediately after it
interacted with the unknown qubit. This is a very
significant practical
advantage of the measurements of intensity $\gamma$. Assuming
that
Alice and Bob chose either bases with frequency 50\%, Eve's overall
information
gain
is:
\beq
\label{eveinfo}
I_{AE}(\gamma) = \half\left[1+\frac{1+\sin\gamma}{2}
\log_2(\frac{1+\sin\gamma}{2})
+\frac{1-\sin\gamma}{2} \log_2(\frac{1-\sin\gamma}{2})\right] \; ,
\eeq
where $\log_2$ is the base two logarithm, to get the information in bits.

A problem associated with these measurements however, is that the error
rate depends on the basis used by Alice and Bob. Therefore, by checking it
independently for the two bases, Alice and Bob may infer that the noise in
their transmission is not produced by a random process.
Moreover they may get some information about Eve's strategy. To avoid
that, Eve should use a random combination of two measurement strategies,
one along the \vbasis axis, as in Eq.~(\ref{mesIgamma}), and one along the
\hbasis axis. It is easy to see that, for this second strategy, the
density matrix of the state received by Bob becomes:
\beq
\label{rhobobp}
\rho_{\rm Bob}^{\prime}=\half\pmatrix{1+\cos\gamma\cos\theta & \sin\theta
\cr
\sin\theta & 1-\cos\gamma\cos\theta} \; ,
\eeq
so that the final density matrix, for the random combination of both
strategies is:
\beq
\label{rhobobav}
\overline{\rho}_{\rm Bob}=\half\pmatrix{1+\eta\cos\theta & \eta\sin\theta
\cr
\eta\sin\theta & 1-\eta\cos\theta} \; ,
\eeq
where $ \eta \equiv \frac{1+ \cos\gamma}{2}$. 
This density matrix is now entirely symmetric with respect to the initial
state. Indeed, if we write the initial density matrix:
$ \rho_{\rm Alice} = \frac{1 + \vec{m}\cdot \vec{\sigma}}{2}$, where
$\vec{m}=(\sin\theta,0,\cos\theta)$ is the Bloch vector representating
$\rho_{\rm Alice}$ on the Poincar\'{e} sphere, the final state is:
$\overline{\rho}_{\rm Bob}= \frac{1 + \eta \vec{m}\cdot \vec{\sigma}}{2}$. 
The effect of Eve's eavesdropping is thus simply to ``shrink'' the Bloch
vector by $\eta$.

\begin{figure}[ht]
\hspace*{1cm}\epsfbox{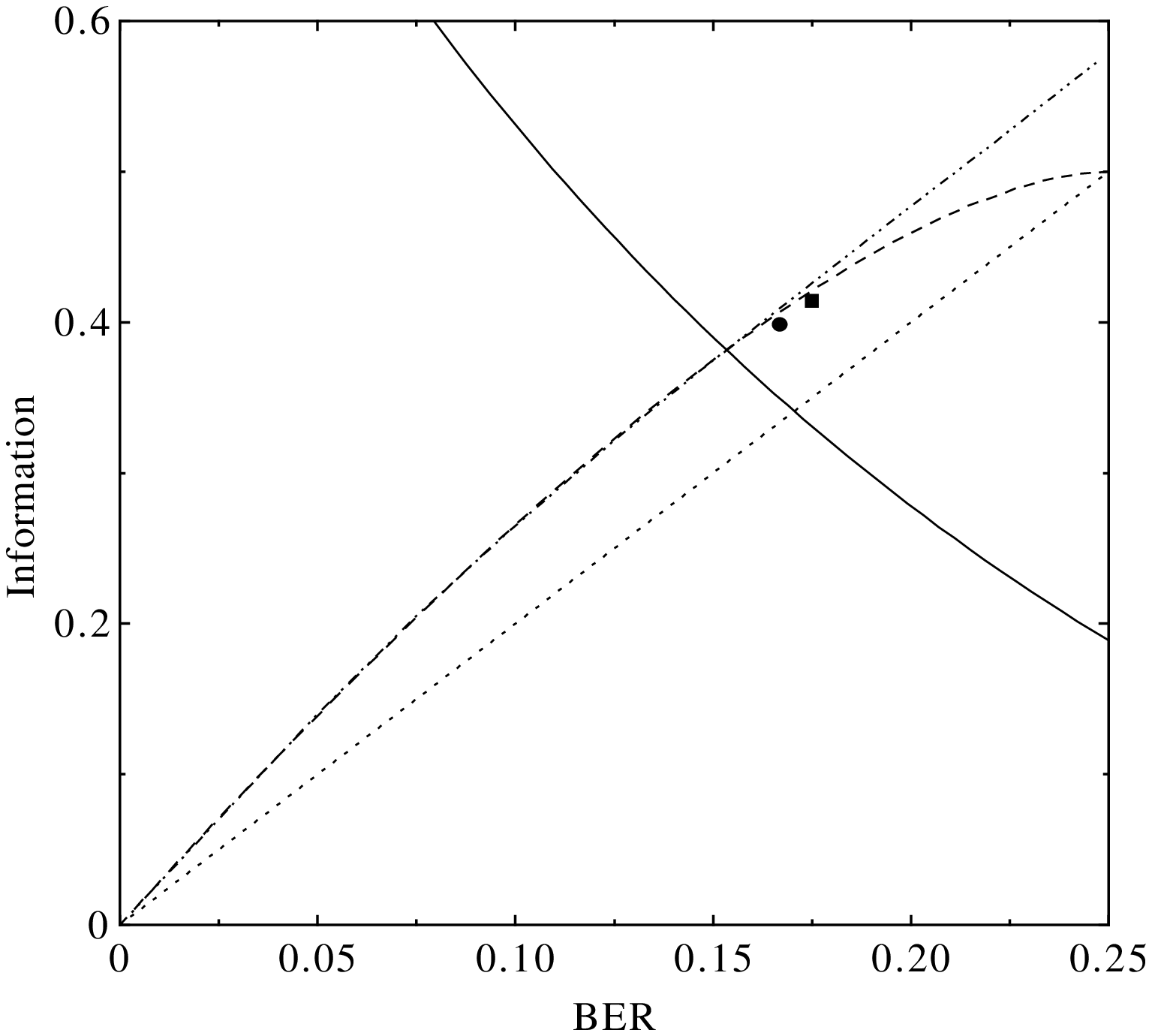}
\caption{}
Information gain versus Bite Error Rate (BER) for various
eavesdroping strategies. The full curve represents the mutual information
between Alice and Bob. The dotted curve is the information available to
Eve for the intercept/resend strategy. The dashed curve is for the
measurement of intensity $\gamma$. The dashed-dotted curve is the optimal
eavesdropping strategy with a 2D probe. The circle is the information
gained by Eve if she uses a UQCM (which requires interaction between three
qubits), while the square is obtained by the PGQCM (which only requires
interaction between two qubits).
\label{info}
\end{figure}

The disturbance of the initial state 
increases with
increasing $\gamma$, as indicated by the fidelity function:
\beq
\F(\gamma)\equiv \bra{\psi(\theta)}\overline{\rho}_{\rm
Bob}\ket{\psi(\theta)}= \frac{3+\cos\gamma}{4}  \; ,
\eeq
or equivalently by the BER, $q(\gamma)$:
\beq
\label{qgamma}
q(\gamma) \equiv 1-\F(\gamma)=\frac{1-\cos(\gamma)}{4} \; .
\eeq
Eve's information gain, Eq.~(\ref{eveinfo}), as a function of the BER,
Eq.~(\ref{qgamma}),  is
depicted on
Fig.~\ref{info}, together with  
Bob's information:
$I_{AB}=1+q \log_2q +(1-q) \log_2(1-q)$. 
Fig.~\ref{info} shows that, above a BER of about 15\%,  Eve may have more
info
than Bob, 
hence no classical error correction and privacy amplification can be
applied. It is interesting to note that this simple strategy is
undistinguishable from the ``optimal'' strategy of~\cite{lutkenhaus}. 

In order to discover the quality of the simple measurements of intensity
$\gamma$, we performed a 
computer optimization of Eve's information $I_{AE}$ for given
BERs, assuming the most general eavesdropping strategy given by
Eqs.~(\ref{Uab}) and~(\ref{alphabeta}). In this case, the density matrix
received by Bob when Alice sends state $\ket{\psi(\theta)}$ is:
\beq
\label{rhobobgen}
\rho_{\rm Bob}=\half\pmatrix{1+\cos\theta \cos 2\alpha & \sin 2\alpha\cos
2\beta + \sin\theta \cos 2\alpha \sin 2\beta \cr
\sin 2\alpha\cos 2\beta + \sin\theta \cos 2\alpha \sin 2\beta &
1-\cos\theta \cos 2\alpha} \; .
\eeq
The density matrix received by Eve is simply obtained by interchanging
$\alpha \longleftrightarrow \beta$. Similarly to the measurement of
intensity $\gamma$, in order to avoid creating an assymmetric state for
Bob, Eve has to use a random combination of strategies. In this case, she
has to choose at random between four possible unitary transformations
similar to Eq.~(\ref{Uab}),  corresponding to
four choices of
symmetry breaking along four directions mutually orthogonal on the
Poincar\'e sphere (e.g. $\up$, $\down$, $\lefta$, $\righta$). The averaged
density matrix for Bob becomes similar to Eq.~(\ref{rhobobav}), with
$\eta= \frac{\cos 2 \alpha (1+\sin 2\beta)}{2}$. Note that, for this more
general strategies, Eve gains information in both bases, \vbasis and
\hbasis. However, since this information is different for the two bases,
it is now preferable for Eve to wait till Alice and Bob announce their
bases publicly, before she measures her probe. The calculation of $I_{AE}$
as a function of $\alpha$ and $\beta$
is more cumbersome but straightforward, and will not be presented here
explicitely. The computer optimization, giving the best $I_{AE}$ at a
given error rate, is represented by
the upper curve of Fig.~\ref{info}. 
The optimum strategy for Eve is a
measurement of
intensity $\gamma$ for low error rates, and remains indistinguishable from
it up to a BER of about 15\%. For example, at the crossing point between
$I_{AB}$ and $I_{AE}$, i.e. for a BER of 0.1534, $I_{AE}^{\gamma}=
0.3816$, while $I_{AE}^{\rm opt}= 0.3820$.  However, above this BER
value, 
Eve can get more information.
In
particular for large BER she can get more than 0.5 bits of information.

\begin{figure}[ht]
\hspace*{3cm}\epsfbox{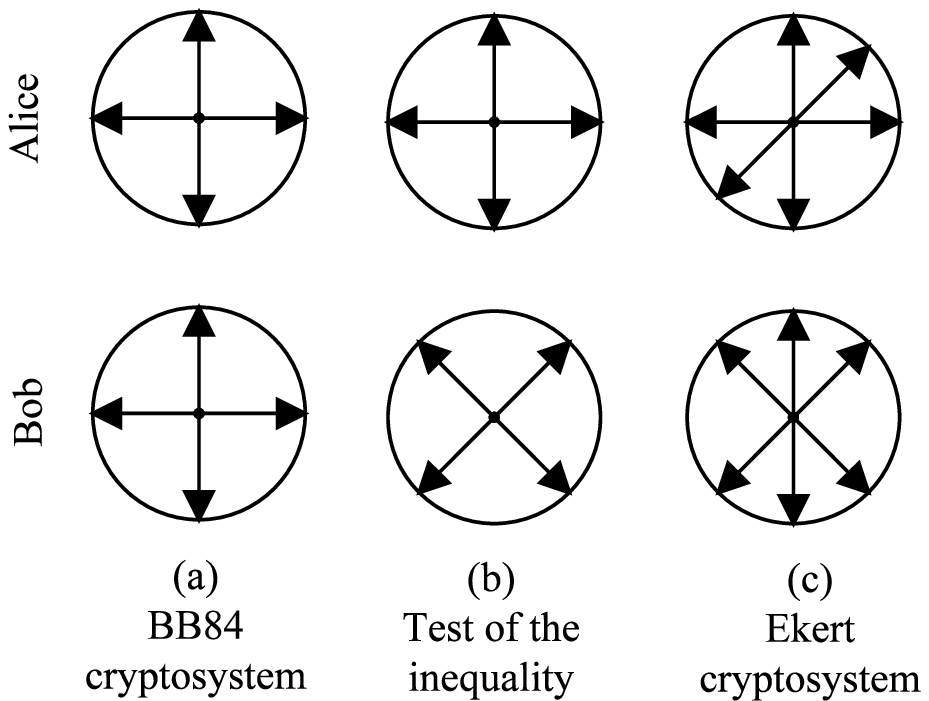}
\caption{The various setups}
To implement the BB84 quantum cryptographic protocol, Alice and Bob use
the same bases to prepare and measure their particles. A representation of
their states on the Poincar\'{e} sphere is shown in~(a). A
similar setup, but with Bob's bases rotated by $45^\circ$, can be used to
test the violation of Bell inequality, as shown in~(b). Finally, in the
Ekert protocol, Alice and Bob may use the violation of Bell inequality to
test for eavesdropping, as shown in~(c). 
\label{setup}
\end{figure}

Let us now look at the case where Alice and Bob wish to test the
inequality to check the integrity of the transmission. Following the above
discussion, and since we are mainly interested in low error rates, we
shall restrict ourselves to the measurement of intensity $\gamma$.
Let us
first analyze the inequality mentionned in the introduction (see
Fig.~\ref{setup}), where the
reference frames of Alice and Bob are rotated by $45^\circ$. Eve's
symmetry axis, i.e. her choice of $\up$ and $\down$ states in
Eq.~\ref{Uab},
may be chosen along any of the axes chosen by Alice and Bob. We can
easily
calculate the value of the parameter $S$ 
that appears in the CHSH version of Bell's inequality~\cite{CHSH69}, to
get:
\beq
S_{AB}=\sqrt{2}(1+ \cos\gamma) \; ,
\eeq
(The same value is also obtained when Eve adopts the symmetrized
strategy). This shows that if the BER is less than $q_{\rm
Bell}=1/2-\sqrt{2}/4 \approx
0.1464$,
Alice and Bob's data violates the inequality. On the other hand, for the
particular setup of Fig.~\ref{setup}(b), for a BER above $q_{\rm Bell}$,
their data does not violate the inequality.

The above discussion only refers to the particular setup of
Fig.~\ref{setup}(b), where the
reference frames of Alice and Bob are at $45^\circ$. A more general
criterion was recently given by the Horodeckis~\cite{Horodecki95}, giving
the necessary and
sufficient condition for a density matrix to violate the CHSH
inequality~\cite{CHSH69} for at
least one particular choice of directions. Applying first this criterion
to our
{\em unsymmetrized} system (i.e. Eve uses only one measurement of
intensity
$\gamma$), we found that Alice and Bob's joint density matrix fulfills
this
criterion for $\gamma < \pi/2$. Therefore, there exists one set of
directions following which Alice and Bob would still violate some CHSH
inequality till an error rate of 25\%.  However,if we now consider {\em
symmetrized}
strategies, which are more likely to be used by Eve in order to avoid
detection, we find that the limit is precisely $q_{\rm Bell}$. In other
words, 
above this limit, the joint density matrix does not violate any CHSH
inequality, and the above choice at $45^\circ$ is optimal for testing for
eavesdropping with CHSH inequalities.

The value of $q_{\rm Bell}\approx 0.1464$ is suspiciously close to the
intersection of the two curves
$I_{AB}$ and $I_{AE}$, at a BER of 0.1534. Therefore, we make the
conjecture that the real optimal strategy, which should be obtained with a
4-D probe, would give this very point\footnote{This conjecture seems to be
validated by very recent results by Fuchs and Peres~\cite{fpnew}, who
analyzed the optimal strategy (obtained with a
4-D probe). Their results show that $I_{AB}=I_{AE}$ for a BER of 0.1464.}.
The violation of the inequality
would then become a interesting measure of the eavesdropping\footnote{This
idea was first expressed by A. Ekert}: if the
inequality is violated, Alice and Bob know that Eve has less mutual
information than themselves, and that they can in principle perform
classical information processing to obtain a secret key. If the inequality
is not violated, then Eve may have more information, and so a secret key
cannot be distilled by classical means. Note that the conjecture is
restricted to symmetrized strategies, so that Alice and Bob still need to
check that the error rate is the same for all four possible states. This
conjecture is not proven yet,
but at least there is no known strategy which contradicts it. 

However, even though above an error rate of  0.1464
Alice and Bob  cannot extract a secret key by classical information
processing, there is still something quantum hidden there. 
Indeed, using the so-called purification of entanglement~\cite{bennett},
Alice and Bob
may still take advantage of their system to extract a secret key. This
procedure, known as quantum privacy amplification or QPA~\cite{qpa},
performs
operations on two pairs of particles at a time, and can extract from the
corrupted
set pairs of entangled states with an arbitrarily high degree of
entanglement, on which Eve is automatically
excluded. In our case, the condition for this algorithm to be effective
reads:
\beq 
\bra{\psi_-} \rho  \ket{\psi_-} > \half \;  ,
\eeq
where $\ket{\psi_-} \equiv \frac{\sqrt{2}}{2}(\ket{\up \down} - \ket{
\down \up})$ is the singlet state, and $\rho$ the joint density matrix of
Alice and Bob. This limit is attained for
$\gamma = \pi/2$, corresponding to an error rate of $25 \%$. 
This proves that
QPA 
is not only ``provably secure'', but is trully more powerful than any
classical privacy amplification algorithm. 

Note that if Eve and Bob both do measurements in the diagonal bases, then
both Eve and Bob
can violate Bell inequality for some values of $\gamma$. Indeed when Alice
and Bob use the setup of Fig.~\ref{setup}(b), and Eve uses 
measurements of intensity $\gamma$, we find
$S_{AE}=2\sqrt{2}\sin\gamma$. Both $S_{AB}$ and
$S_{AE}$ 
are therefore larger than 2 for $\gamma$ around $\pi/3$.

\section{Quantum cloning and broadcasting of quantum
information}\label{Qclone}
In this Section we consider the question of (imperfect) quantum cloning
and
its possible application for eavedropping and quantum information
broadcasting.
First, we consider ``classical'' quantum cloning machines in which the
only
relevant quantum degrees of freedom are two qubits, one to be copied and
one to receive the copy. Hence, we can use the same general frame as in
the
previous Section~\ref{Mgamma}, that is the unitary transformation
defined
by Eq.~(\ref{Uab}) with the parameters $\alpha$ and $\beta$ as defined by
Eq.~(\ref{alphabeta}). In this case, what we want is the two density
matrices $\rho_{\rm Bob}$ and $\rho_{\rm Eve}$ to be equal, and as close
as
possible to the original $\rho_{\rm Alice}$. For the moment, the
distinction between Bob  and Eve is not relevant, as both now have the
same kind of copy of the original state. However, since we later want to
use this copying machine for eavesdropping, we keep the terminology. From
Eq.~(\ref{rhobobgen}),
the two matrices are equal when $\alpha=\beta$. 
In order to give a quantitative measure of the quality of 
our cloning machine, we use the mean fidelity with respect to the
perturbed original
qubit\footnote{This criterion is certainly not the only one that can be
adopted, but it gives simple and reasonable results}:
\beq
\bar\F = \frac{1}{2 \pi} \int_0^{2\pi} \bra{\psi(\theta)}\rho_{\rm
Bob}\ket{\psi(\theta)} d \theta \; .
\eeq
It is now straightforward to find the maximum value: 
\beq
\bar\F_{\rm opt}  = \frac{8+3\sqrt{3}}{16} \approx 0.825 \; ,
\label{MeanF}
\eeq
which is reached for $\alpha=\beta=\pi/12$.
Interestingly, it is also possible to relax the equality condition, and
simply optimize the sum of the fidelities of the two copies.  The result
is the same as Eq.~(\ref{MeanF}). 
We like to call this cloning machine a  Pretty Good Quantum Copying
Machine (PGQCM),
for reasons described below.

\begin{figure}[ht]
\hspace*{2cm}\epsfbox{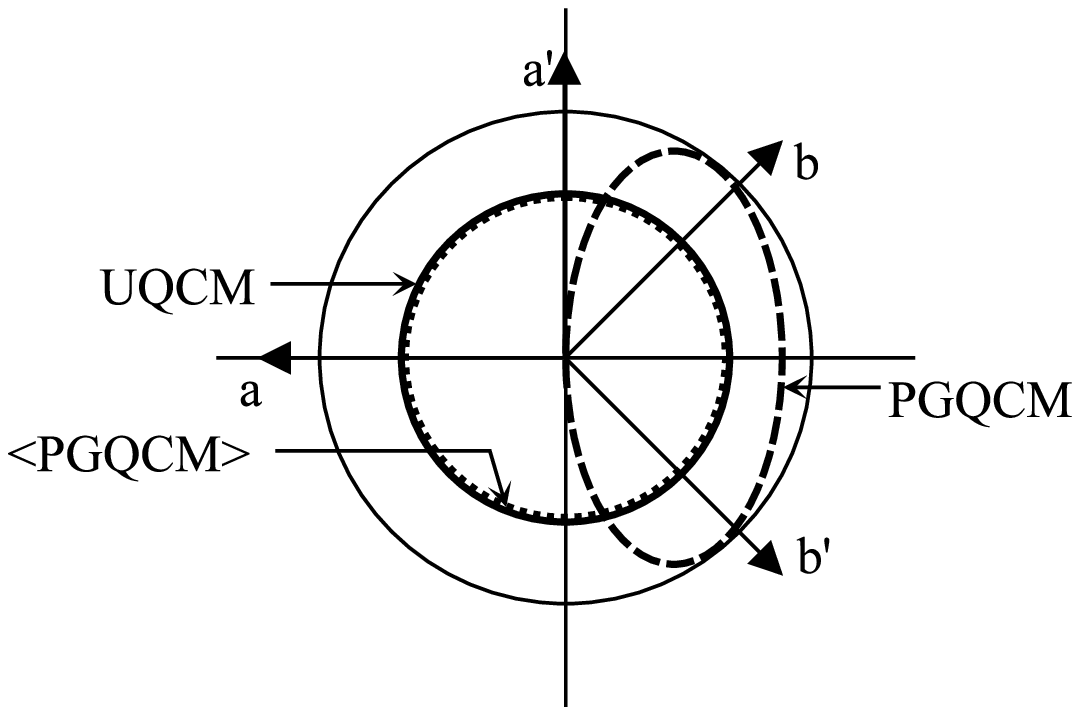}
\caption{}
Bloch vector representation of the states produced by the UQCM (full
line), the 
PGQCM (dashes) and the symmetrized PGQCM (dots)
corresponding to
input state represented by a large circle on the Poincar\'e sphere. The
directions $\bf a$, $\bf a^\prime$, $\bf b$ and $\bf b^\prime$ are used
for testing the inequality. We see that the choice of the symmetry
breaking axis for the PGQCM influences the possible violation of the
inequality.
\label{PGQCM}
\end{figure}

Fig.~\ref{PGQCM} displays the Bloch vector corresponding to the copied
states for original
states corresponding
to a large circle on the Poincar\'e sphere. Note that these  are
highly non symmetric. However, the symmetry can be restored if the cloning
machine uses at random four similar unitary transformation, corresponding
to
four choices of
symmetry breaking along four directions mutually orthogonal (on the
Poincar\'e sphere),similarly to the previous eavesdropping strategy. 
This amounts to add
an additional degree of freedom, but this one can be classical, hence
easy to produce in the lab. The Bloch vector with such
a copy machine is the same as given in Eq.~(\ref{MeanF}), and is also
shown in Fig.~\ref{PGQCM}. In this case the Bloch vectors of the copies
are the same, $\vec{m}_{\rm copy}$, and are simply related to the Bloch
vector of the  original qubit $\vec{m}_{\rm ini}$ by $\vec{m}_{\rm copy} =
\eta \vec{m}_{\rm ini}$, where $\eta= 2 \bar\F_{\rm opt} -1$.

Recently Bu\v{z}ek and Hillery~\cite{buzek} have introduced another
cloning
machine, the
UQCM (Universal Quantum Cloning Machine). This is a truly quantum
mechanical
machine in the sense that in addition to the minimum two qubits, the
machine
itself has quantum degrees of freedom, although these can be described by
a
2-D Hilbert space with basis vectors $M_\up$ and $M_\down$. Their
machine
is described by the following unitary transformation:
\beqa
\ket\up\otimes\ket 0 \otimes\ket M_o &\rightarrow&
\sqrt{\frac{2}{3}}\ket{\up\up} \otimes \ket {M_\up} +
\sqrt{\frac{1}{6}}(\ket{\up\down}+\ket{\down\up})\otimes \ket {M_\down}
\\ \nonumber
\ket\up\otimes\ket 0\otimes\ket M_o &\rightarrow&
\sqrt{\frac{2}{3}}\ket{\down\down}\otimes\ket {M_\down} +
\sqrt{\frac{1}{6}}(\ket{\up\down}+\ket{\down\up})\otimes\ket {M_\up}
\label{UQCM}
\eeqa
As for the PGQCM, the density matrices  of both copies are equal.
Moreover, this machine is already symmetric, with fidelity $\bar\F_{\rm
uni}$ independent of the initial state: 
\beq
\bar\F_{\rm uni}= \frac{5}{6}
\approx 0.833 \; . 
\eeq
The corresponding
Bloch vector is also plotted in Fig.~\ref{PGQCM}. 
Note that the UQCM has a slightly higher mean fidelity than the PGQCM
(about 1\% larger)
at the cost of the complication due to the additional quantum degree of
freedom, hence our vocabulary of ``pretty good'' for our PGQCM.

Eve may use either of these copying machines to tape into Alice and Bob
quantum communication channel. Here, contrary to the case of measurement
of
intensity $\gamma$ studied in Section~\ref{Mgamma}, Eve gains to wait
until Alice and Bob reveal their bases, before measuring her copy.
Moreover, for the case of the UQCM, since the state of the machine itself
after the interaction depends on the initial state, Eve may use this extra
information contained in the machine. 
Her information gain for given BERs is added on Fig.~\ref{info}, both for
the PGQCM and the UQCM. We see that even
the UQCM provide less information than the optimal strategy using only a
2-D probe.

Finally, one may also use either cloning machines to send copies to two
different Bobs, let say $\rm Bob_1$ and $\rm Bob_2$. Each Bob can then
independently
measure his qubits. We find out that, when we use the assymetric PGQCM,
both Bobs violate the inequality with
respect to the same
Alice
data. It is clear from Fig.~\ref{PGQCM} that this is dependent on the
choice of directions $a$, $a^\prime$, $b$ and $b^\prime$. For the
symmetrized version,  $S_{AB_1}=S_{AB_2}=2\sqrt{2}( 2 \bar\F -1)  < 2$, so
that there is no violation. It
should also be noted that the results of $\rm Bob_1$ and $\rm Bob_2$ are
not
independent,
because the cloning entangles the original and the clone qubits.

\section{Conclusion}\label{concl}
The best possible eavesdropping strategy remains to be explored. However,
we have
seen that even using only one qubit as a probe, i.e. using techniques
already under
development in several labs,
Eve can do better than the standard intercept/resend strategy. 
We have found the optimal strategy for such probes, and shown that for low
BER, it is indistinguishable from a simple strategy termed measurement of
intensity $\gamma$. This strategy puts a limit of about
15\% on the BER above which Eve may possess more information  than Bob,
which means that classical information processing 
can not be used to distill a secret key. However, for the same
eavesdropping strategy, quantum information processing, and more precisely
the QPA may still be used, up to a BER of 25\%-$\epsilon$,
for any $\epsilon >0$. Hence, 
quantum privacy amplification not only enables to distribute the key in a
provably secure fashion, but is also
intrinsically more powerful than its classical analog. 
Amusingly, using the strategy described in Section~\ref{Mgamma}, Eve can
extract enough information from the quantum channel to
violate the inequality, while simultaneously perturbing the quantum
channel so little that Bob could still violate the inequality (with
respect to the same data of Alice).

Perfect quantum cloning is impossible. However, an arbitrary qubit can be
cloned
in such a way that both the perturbed original qubit and the cloned qubit
both
have a fidelity above 80\%, independently of the initial state to be
copied. 
Such cloning machine can be used to broadcast
quantum information, for quantum cryptography purposes or for the fun of
multi-violation of Bell's inequalities. The Pretty Good Quantum Cloning
machine
presented in Section~\ref{Qclone}, while about 1\% less efficient than the
Universal Quantum Copy Machine, is close to feasible with todays
technology. Both this PGQCM and the measurement of intensity $\gamma$ are
interesting examples of potential uses for the simplest quantum gates,
with only two interacting qubits, which are now under development in
various laboratories.

\section*{Acknowledgments}
We would like to thank A. Peres for a very fruitful discussion, and  S.
Popescu and T. Mor for useful comments.  We gratefully acknowledge
financial support from the Swiss Fonds National
de Recherche Scientifique.

\newpage

\section*{Figures}

\end{document}